\documentclass[prl,twocolumn,showpacs,amsmath,amssymb]{revtex4}
\usepackage{graphicx}% Include figure files
\usepackage{dcolumn}% Align table columns on decimal point
\usepackage{bm}% bold math
\usepackage{amssymb}% bold math

\begin{document}

%\begin{verbatim}

%\preprint{APS/123-QED}

\title{Neutron scattering study of novel magnetic order in Na$_{0.5}$CoO$_{2}$}
% Force line breaks with \\

\author{G.~Ga\v{s}parovi\'{c},$^1$ R.~A.~Ott,$^1$
J.-H.~Cho,$^{1,3}$ F.~C.~Chou,$^{2}$ Y.~Chu,$^{1,2}$
J.~W.~Lynn,$^4$ and Y.~S.~Lee$^{1,2}$}
\affiliation{%
\begin{center}
\mbox{$^1$Department of Physics, Massachusetts Institute of
Technology, Cambridge, MA 02139}\\
\mbox{$^2$Center for Materials Science and Engineering,
Massachusetts Institute of Technology, Cambridge, MA 02139}\\
\mbox{$^3$RCDAMP and Department of Physics, Pusan National University,
Busan 609-735, Korea}\\
\mbox{$^4$NIST Center for Neutron Research, Gaithersburg, Maryland
20899}
\end{center}
}%
\date{\today}

\begin{abstract}

%PACS numbers may be entered using the \verb+\pacs{#1}+ command.

We report polarized and unpolarized neutron scattering
measurements of the magnetic order in single crystals of
Na$_{0.5}$CoO$_{2}$. Our data indicate that below $T_N=88$~K the
spins form a novel antiferromagnetic pattern within the CoO$_2$
planes, consisting of alternating rows of ordered and non-ordered
Co ions. The domains of magnetic order are closely coupled to the
domains of Na ion order, consistent with such a two-fold symmetric
spin arrangement. Magnetoresistance and anisotropic susceptibility
measurements further support this model for the electronic ground
state.

\end{abstract}
%\keywords{Na$_{x}$CoO$_{2}$, neutron scattering, 
%strongly correlated electrons, superconductivity}
%Use showkeys class option if keyword
\maketitle

The layered cobaltates Na$_x$CoO$_2$ have attracted much attention
due to their unusual thermoelectric
properties~\cite{terasaki,wang} and because of the recent
discovery of superconductivity in the hydrated
composition~\cite{takada1}.  The Co ions in these compounds form a
hexagonal lattice, and the average valence can be changed by
varying the Na concentration $x$.  The Co$^{4+}$ ions can formally
be regarded as magnetic ions with $S=\frac{1}{2}$ on a frustrated
low-dimensional lattice.  It is believed that strong electronic
correlations play an important role in the physics, and, as a
function of the Na concentration, a rich phase diagram has been
reported~\cite{foo1}. For $x<\frac{1}{2}$, the material is a
paramagnetic metal, while for $x>\frac{1}{2}$ an unusual
``Curie-Weiss metallic'' phase is observed.  At $x=\frac{1}{2}$, a
unique state is realized: the Na ions are chemically ordered to
form zig-zag chains in an orthorhombic superstructure with
two-fold symmetry~\cite{huang1,zandbergen1}, and it has been
speculated that this leads to Co$^{3.5+\delta}$/Co$^{3.5-\delta}$
charge order within each CoO$_{2}$
layer~\cite{foo1,wang_nl,timusk,pickett,phillips}. The ground
state is believed to be a magnetically ordered insulator;
resistivity, Hall coefficient, thermal transport~\cite{foo1}, and
angular magnetoresistance oscillation measurements~\cite{balicas}
are all consistent with such a two-fold symmetric ground state.

In order to further investigate the electronic ground state of the
half-doped CoO$_2$ plane, we have performed neutron scattering,
susceptibility, and transport studies on single crystal samples.
The combination of polarized and unpolarized neutron scattering
data allow us to determine the ordered spin direction and
arrangement.  We find that the ground state is well described by
an ordered array of stripes of antiferromagnetic spins interleaved
with stripes of non-ordered Co ions.  Our transport and
anisotropic susceptibility data further support such a model.

%====================================================================
\begin{figure}
\includegraphics[angle=0,width=0.44\textwidth]{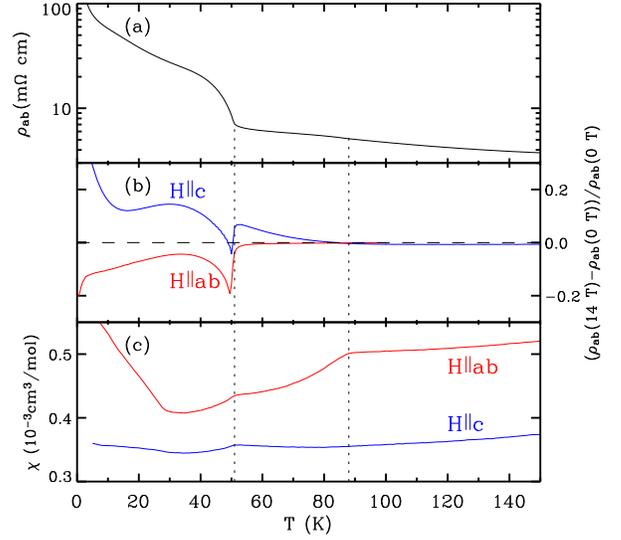}
\vspace{-1mm} \caption{Bulk physical properties measured using
Na$_{0.5}$CoO$_{2}$ single crystals. (a) In-plane resistivity
$\rho_{ab}$.  (b) Fractional change of $\rho_{ab}$ measured in a
magnetic field of 14 T with $H\parallel c$ and $H \parallel ab$.
(c) Anisotropic magnetic susceptibility in a field of 1~T.  The
vertical dashed lines denote the temperatures of $T=51$~K and
$T_N=88$~K. } \vspace{-4mm}
\end{figure}
%====================================================================

The samples used in this study were prepared by electrochemically
deintercalating a floating-zone grown Na$_{x}$CoO$_{2}$ crystal to
yield a final concentration of $x=0.5$~\cite{chou2,chou1}.
Figure~1 shows measurements of the bulk properties of the
resulting Na$_{0.5}$CoO$_2$ crystals. In Fig.~1(a), we plot the
in-plane resistivity ($\rho_{ab}$) measured using the conventional
four-probe method on a crystal with dimensions $1\times 0.5\times
0.05$ mm$^3$ in zero applied field. The resistivity dramatically
increases below 51~K, consistent with previously published
data~\cite{foo1}.  The application of a 14 T magnetic field
affects $\rho_{ab}$ in different ways depending on the field
orientation, as shown in Fig.~1(b). For $H\parallel {ab}$, a
negative magnetoresistance is observed below $\sim 51$~K,
consistent with previous work by Balicas {\it et
al.}~\cite{balicas}. The kink in the data near 50 K is caused by a
shift in the onset temperature of the resistivity upturn.  A new
observation is that, for $H\parallel c$, the data exhibit a
positive magnetoresistance below 88 K.  As we discuss further
below, this behavior coincides with the magnetic phase transition
at $T_N=88$~K and is consistent with our proposed ground state.
The magnetic susceptibility (Fig.~1(c)) is anisotropic and
exhibits kinks at $T=88~K$ and 51~K~\cite{chou_mag,yokoi}.

Our neutron scattering experiments were conducted at the NIST
Center for Neutron Research using the BT9 and BT2 triple-axis
spectrometers. The incident energy was fixed at 14.7 meV, and the
collimations were
40$^{\prime}$-40$^{\prime}$-sample-60$^{\prime}$-open.  Pyrolytic
graphite filters were placed in the beam before the sample to
reduce higher order neutron contamination.  For the polarized
measurements, Heusler crystals were used for the monochromator and
analyzer, and polarization guide fields were created at the sample
position using Helmholtz coils (a flipping ratio of $\sim 21$ was
achieved). The crystal studied was cylindrically shaped, with mass
2.41~g and a mosaic of 1 to 2$^{\circ}$ (full-width half maximum),
depending on the orientation.  The temperature was controlled
using a $^{4}$He closed cycle refrigerator. We label reciprocal
space peaks using the hexagonal $P6_{3}/mmc$ space group, with low
temperature lattice parameters $a=2.816$~\AA, and $c=11.05$~\AA.
As mentioned previously, in Na$_{0.5}$CoO$_{2}$, Na ion ordering
leads to an orthorhombic supercell ($a_o=\sqrt{3}a$ and $b_o=2a$)
in real space, which produces nuclear superlattice reflections in
reciprocal space~\cite{zandbergen1}.

%====================================================================
\begin{figure}
\includegraphics[angle=0,width=0.47\textwidth]{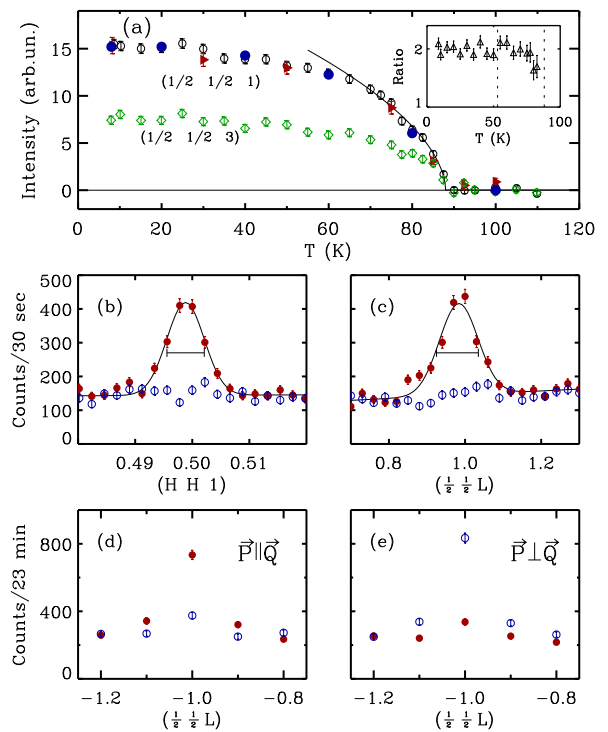}
\caption{(a) Temperature dependence of the intensities of the
$(\frac{1}{2}~\frac{1}{2}~1)$ ($\circ$) and
$(\frac{1}{2}~\frac{1}{2}~3)$ ($\diamondsuit$) magnetic Bragg
peaks measured by neutron diffraction (a background was
subtracted). The filled triangles $(\blacktriangleright)$ denote
the integrated intensities obtained from $\theta$-scans, and the
filled circles $(\bullet)$ denote the intensities measured using
polarized neutrons.  The solid line corresponds to $I\propto
\langle M \rangle^{2} \propto (T_N-T)^{2\beta}$ with $T_N=88$~K
and $\beta=0.28(1)$. Inset: the ratio of the intensity of
$(\frac{1}{2}~\frac{1}{2}~1)$ to that of
$(\frac{1}{2}~\frac{1}{2}~3)$. (b) $(HH1)$ scan and (c)
$(\frac{1}{2}~\frac{1}{2}~L)$ scan through the magnetic
$\vec{Q}=(\frac{1}{2}~\frac{1}{2}~1)$ peak at $T=8$~K ($\bullet$)
and $T=120$~K ($\circ$).  The solid lines are Gaussians\ fits, and
the horizontal bars indicate the experimental resolution. The
lower panels show polarized neutron diffraction with (d) $\vec{P}
\parallel \vec{Q}$ and (e) $\vec{P}
\perp \vec{Q}$, for spin-flip ($\bullet$) and non-spin-flip
($\circ$) channels at $T=8$~K.} \vspace{-3mm}
\end{figure}
%====================================================================

Figure 2 shows unpolarized neutron diffraction evidence for
magnetic order below $T_N=88$~K.  At low temperatures, new peaks
arise which are distinct from the Na superlattice reflections.  In
the $(HHL)$-zone, we observe such peaks at
$\vec{Q}=(\frac{1}{2}~\frac{1}{2}~odd)$.  The temperature
dependence of the peak intensities of the
$(\frac{1}{2}~\frac{1}{2}~1)$ and $(\frac{1}{2}~\frac{1}{2}~3)$
peaks are shown in Figure 2(a). Also plotted are the integrated
intensities of the $(\frac{1}{2}~\frac{1}{2}~1)$ reflection which
match the $T$-dependence of the peak intensity.  Figures 2(b) and
2(c) show reciprocal-space scans through the
$(\frac{1}{2}~\frac{1}{2}~1)$ peak along the $(HH0)$ and $(00L)$
directions, respectively, at $T=8$~K and $T=120$~K.  The low
temperature peak widths are resolution-limited, indicating
long-range order ($>50$ \AA).

In order to confirm that these peaks originate from magnetic
order, we performed polarized neutron measurements. This is the
clearest way to distinguish magnetic peaks from weak nuclear
superlattice peaks. In general, the magnetic diffraction cross
section is given by \mbox{$(\frac{d\sigma}{d\Omega}) \sim
\sum_{i,f} \vert\langle f \vert\sum_{l}
e^{i\vec{Q}\cdot\vec{r}_{l}}U_{l}^{S_{i}S_{f}}\vert
i\rangle\vert^{2}$}, where $i$ and $f$ denote the initial and
final states of the system. $U_{l}^{S_{i}S_{f}}~\sim F(\vec{Q})
\langle S_{f} \vert \vec{M}_{\perp l}\cdot\vec{\sigma}\vert S_{i}
\rangle$ is the scattering amplitude for neutron spin state
$S_{i}$ to $S_{f}$ at atomic site $l$, where $F(\vec{Q})$ is the
magnetic form factor, $\vec{M}_{\perp}$ is the component of the
magnetic moment perpendicular to the scattering wavevector
$\vec{Q}$, and $\vec{\sigma}$ is the spin operator.

We have measured both the spin-flip and non-spin flip
cross-sections with neutron polarizations both parallel and
perpendicular to the wavevector $\vec{Q}$.  Figure~2(d) shows the
scattering measured at $T=8$~K with $\vec{P}
\parallel \vec{Q}$ at the $(\frac{1}{2}~\frac{1}{2}~1)$ peak in
the $(HHL)$-zone.  In this geometry, all magnetic scattering will
occur in the spin-flip channel, and nuclear scattering will be
non-spin-flip.  The data show that the scattering is entirely
spin-flip.  The small peak in the non-spin-flip channel is
consistent with the background measured at high-temperature (not
shown), taking into account the instrumental flipping ratio.  The
temperature dependence of this scattering (scaled to match the
unpolarized data at low-$T$) is denoted by the filled circles in
Figure~2(a). We see that the temperature dependence is identical.
Hence, this proves that the scattering below 88~K at
$(\frac{1}{2}~\frac{1}{2}~1)$ is entirely magnetic.

Neutron polarization analysis is also an extremely useful method
to deduce the direction of the ordered moment.  In Figure 2(e), we
plot the scattering data measured with $\vec{P} \perp \vec{Q}$;
that is, the $(HHL)$-zone of the sample is horizontal and the
neutron polarization direction is vertical.  In this geometry,
magnetic scattering can occur in the non-spin-flip channel if the
ordered moment has a component parallel to the polarization
direction.  A peak in the spin-flip channel would arise from
ordered moments with a component in the $(HHL)$-plane. Here, the
signal predominately occurs in the non-spin-flip channel.
Therefore, the direction of the ordered moments lies within the
$ab$ plane, and the ordered moment along the $c$-axis must be
small (or zero). The error bars on our data allow us to conclude
that the size of the $c$-axis component which contributes to the
scattering at $(\frac{1}{2}~\frac{1}{2}~1)$ is less than $\sim
0.04 \mu_B$. To further specify the direction of the moments
within the plane, the twin domain distribution must be taken into
account, as we discuss below. The temperature dependence of the
polarized data indicates that the $c$-axis component remains small
at all measured temperatures below $T_N=88$~K.

%====================================================================
\begin{figure}
\includegraphics[angle=0,width=0.47\textwidth]{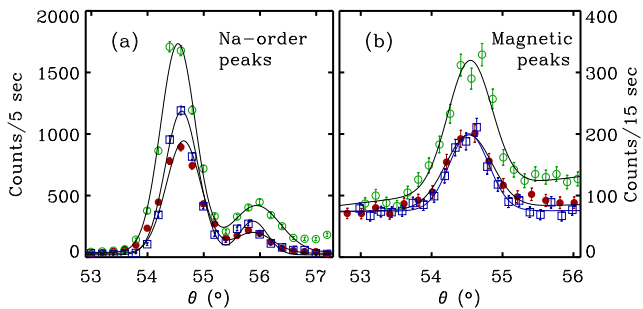}
\caption{(a) Neutron diffraction peaks measured in the
$(HK0)$-zone at $T=8$~K.  The three sets of data correspond to
peaks from three different twin domains due to Na ion ordering:
$(\frac{1}{2}~\frac{1}{2}~0)$ $(\bullet)$,
$(1~\bar{\frac{1}{2}}~0)$ $(\circ)$, and $(\bar{\frac{1}{2}}~1~0)$
$(\square)$ (the angle $\theta$ was offset by $60^{\circ}$ for two
of the peaks).  The weak peak at $\sim 56^{\circ}$ is due to a
small crystallite. (b) Magnetic Bragg peaks measured in scattering
zones tilted from the $(HK0)$-zone by $14.3^{\circ}$ in order to
reach the positions $(\frac{1}{2}~\frac{1}{2}~1)$ $(\bullet)$,
$(1~\bar{\frac{1}{2}}~1)$ $(\circ)$, and $(\bar{\frac{1}{2}}~1~1)$
$(\square)$. The solid lines correspond to Gaussian fits. }
\vspace{-3mm}
\end{figure}
%====================================================================

%====================================================================
\begin{table}
\begin{center}
\begin{tabular}{@{}l@{}r@{}}
\begin{minipage}[t]{125pt}
(a) Nuclear Bragg peaks
\begin{tabular}{ c @{\ } r@{}l@{}l @{\ } r@{}l }
\hline\hline \mbox{$\mathbf{Q}$}&
\multicolumn{3}{c}{$I^{\rm{nucl}}_{\rm{exp}}$} &
\multicolumn{2}{c}{$I^{\rm{nucl}}_{\rm{calc}}$} \\
\hline
$(1~0~0)$ &\ \ \  561&&(12) &\ \ \  603&  \\
$(0~0~4)$ & 624&&(27) & 568&  \\
$(\frac{1}{2}~\frac{1}{2}~0)$ & 104&&(9) & 96& \\
$(\frac{1}{2}~0~2)$ & 9.2&&(0.4) & 10&.0 \\
\hline\hline
\end{tabular}
\end{minipage}
&
\begin{minipage}[t]{125pt}
(b) Magnetic Bragg peaks
\begin{tabular}{ c @{\ } r@{}l @{\ } r@{}l }
\hline\hline \mbox{$\mathbf{Q}$}& \multicolumn{2}{l}{\ \
$I^{\rm{magn}}_{\rm{exp}}$} &
\multicolumn{2}{l}{\ $I^{\rm{magn}}_{\rm{calc}}$} \\
\hline
$(\frac{1}{2}~\frac{1}{2}~1)$ &\ \ \ 3&.9(0.3) &\ \ \  3&.4\\
$(\frac{1}{2}~\frac{1}{2}~3)$ & 2&.2(0.1) & 2&.3\\
$(\frac{1}{2}~\frac{1}{2}~5)$ & 0&.6(0.3) & 1&.2\\
$(\frac{1}{2}~0~1)$ & 8&.1(4.4) & 4&.4\\
$(\frac{1}{2}~0~3)$ & 3&.9(1.6) & 3&.8\\
\hline\hline
\end{tabular}
\end{minipage}
\end{tabular}
\end{center}
\vspace{-2mm} \caption{Observed and calculated integrated
intensities of selected nuclear and magnetic Bragg peaks at
$T=8$~K.  The large error bars for the $(\frac{1}{2}~0~odd)$
magnetic peaks are due to the presence of nuclear superlattice
peaks at the same positions. For these peaks, the intensities at
$T=120$~K were subtracted from the intensities at $T=8$~K.  The
calculated intensities are based on a model which includes the
twin domain distribution, an isotropic Co$^{4+}$ magnetic form
factor, and the instrumental resolution function.} \vspace{-3mm}
\end{table}
%====================================================================

The refinement of the magnetic structure is complicated by the
presence of multiple twin domains (structural and magnetic). These
domains, in principle, can have different populations.  To
characterize the structural domains, diffraction measurements were
taken with the sample aligned in the $(HK0)$-zone. Since the Na
ion order reduces the six-fold hexagonal symmetry to two-fold,
there are three possible Na-order domains whose relative
orientations differ by a rotation of $60^\circ$ about the
$c$-axis. In Fig.~3(a), we show $\theta$-scans through peaks at
$\vec{Q}=(\frac{1}{2}~\frac{1}{2}~0)$, $(1~\bar{\frac{1}{2}}~0)$,
and $(\bar{\frac{1}{2}}~1~0)$.  Each peak is comprised of
scattering from the Na superlattice of two different structural
twin domains.  The measured intensities of the peaks are
proportional to the populations of the twin domains which
contribute to that peak.  (We note that the long axis of our
cylindrically shaped sample lies in the scattering plane in this
geometry.  Hence, part of the intensity variation is related to
neutron absorption.)  The Na order peaks are resolution limited in
all three crystallographic directions, indicating that they are
macroscopic in size.  Also, the relative intensities of these
superlattice peaks compared to the fundamental reflections, shown
in Table I, indicate that the Na ordering occurs in essentially
100\% of our sample.

In conjunction with characterizing the structural twins, we have
also performed scans through the magnetic positions:
$(\frac{1}{2}~\frac{1}{2}~1)$, $(1~\bar{\frac{1}{2}}~1)$, and
$(\bar{\frac{1}{2}}~1~1)$.  These scans, shown in Fig.~3(b), were
performed by first aligning a $(\frac{1}{2}~\frac{1}{2}~0)$ peak,
then tilting the sample by 14.3$^\circ$ to bring the
$(\frac{1}{2}~\frac{1}{2}~1)$ position into the horizontal
scattering plane.  We find that the intensities of these magnetic
peaks are proportional to the intensities of the corresponding Na
order peaks.  This suggests that each structural twin domain
corresponds to a magnetic twin domain. That is, the arrangement of
ordered magnetic moments is closely coupled to the arrangement of
the Na ions.  This is a natural expectation due to the symmetry of
both arrangements and now has experimental support.

%====================================================================
\begin{figure}
\includegraphics[angle=0,width=0.40\textwidth]{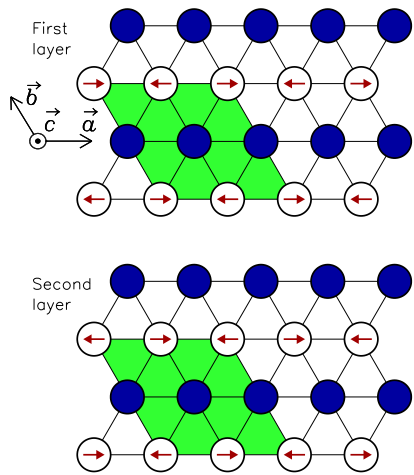}
\vspace{-4mm} \caption{Model of the spin arrangement in
Na$_{0.5}$CoO$_{2}$. The two CoO$_2$ layers in the magnetic unit
cell are depicted, and, for clarity, only Co ions are shown.  The
solid circles represent ``non-ordered'' Co ions, and the hollow
circles represent magnetically ordered Co ions where the arrows
indicate the directions of the magnetic moments. The shaded
parallelogram outlines the magnetic unit cell.} \vspace{-4mm}
\end{figure}
%====================================================================

The above information on the domain distribution and the moment
direction can be combined with measurements of intensities of
several nuclear and magnetic Bragg peaks to determine the magnetic
structure.  Our data are well described by the model shown in
Figure~4.  Within each CoO$_{2}$ layer, for a single magnetic twin
domain, the Co ions are arranged in alternating rows (or stripes)
of magnetically ordered and ``non-ordered'' ions. For the magnetic
ions, the nearest neighbor spins are coupled
antiferromagnetically, both along the row and between rows.
Between CoO$_2$ planes, the rows of magnetic Co ions are stacked
directly on top of each other, and the nearest neighbor interplane
coupling is also antiferromagnetic.  The directions of the ordered
moments are parallel to the rows.  The error bars on our data
allow us to put restrictions on the deviations of the ordered
moments away from this direction to be less than 10$^\circ$ out of
the $ab$ plane. However, within the $ab$ plane, our data are less
restrictive, and the moments may be oriented up to 30$^\circ$
away from the stripe direction.  The low temperature ($T=8$ K)
static magnetic moment is 0.25(2) $\mu_{B}$ per magnetic Co ion.
The size of the observed moment is smaller than that expected for
$S=1/2$, which may suggest the presence of quantum fluctuations or
deviations from a local moment picture.  The order that we observe
differs somewhat from the models proposed by Yokoi {\it et
al.}~\cite{yokoi}, though we agree on the fundamental periodicity.

The structure proposed in Figure 4 is the simplest model which is
consistent with our data, assuming a local moment description of
the magnetic order.  It is consistent with having rows of
Co$^{4+}$ ions which are magnetic (in the $S=1/2$ low spin state)
alternating with rows of Co$^{3+}$ ions (in the $S=0$ spin state)
which are non-magnetic, as suggested previously~\cite{phillips}.
However, we emphasize that our neutron data do not directly probe
the valence of the Co ions, only their magnetic moments.  For
example, the rows of non-ordered ions may correspond to Co ions
with fluctuating moments which do not contribute to the long-range
antiferromagnetism.  The valence difference between the two
distinct Co sites may, in fact, be small.  Recent NQR studies
suggest that unit-valence charge disproportionation does not occur
at $T_N$~\cite{alloul}. It is possible that some degree of valence
disproportionation exists already at temperatures above $T_N =
88$~K, and it is primarily the spin components which order at
$T_N$~\cite{phillips}.  A delocalized moment model may also
describe the data, so long as the periodicity and spin orientation
are the same as discussed.

The anisotropic magnetic susceptibility measurements shown in
Figure 1(c) are consistent with our model.  A clear drop is seen
at $T_N=88$~K for $H\parallel ab$, whereas no kink is visible for
$H\parallel c$, consistent with the observation that there is
essentially no $c$-axis component to the ordered moment below
$T_N$.  The susceptibility data also show an almost isotropic drop
below $T=51$~K, coincident with the upturn in $\rho_{ab}(T)$. A
previous $\mu$SR study suggested that a spin-reorientation
transition occurs near this temperature~\cite{mendels}. We find
that there is no indication of this transition in the magnetic
neutron diffraction. Specifically, the ratio of intensities of the
$(\frac{1}{2}~\frac{1}{2}~1)$ and $(\frac{1}{2}~\frac{1}{2}~3)$
Bragg peaks should change if there were a significant
spin-reorientation below $T=51$~K, but the inset of Fig.~2(a)
shows that the ratio is constant. This is consistent with the
polarized neutron results and implies any spin reorientation must
be small ($<10^{\circ}$ out of the $ab$ plane). Further, this
suggests that the drop in susceptibility near $T=51$~K is not
caused by the spins which participate in the magnetic order below
$T_N=88$~K.  Hence, this feature may be related to a gap
developing in the excitation spectrum of the ``non-ordered'' Co
ions. Finally, the positive magnetoresistance observed below
$T_N=88$~K in Figure 1(b) can be understood in terms of the
orbital motion of charge carriers within this stripe-like ground
state.  The increased resistivity for $H\parallel c$ may be caused
by enhanced scattering for charge motion along one-dimensional
chains due to the Lorentz force.  Understanding the interplay
between spin order and charge motion in this correlated electron
system remains an important issue for further study.

We thank P. A. Lee, P. Phillips, and Q. Huang for fruitful
discussions. The work at MIT was supported by the Department of
Energy under grant number DE-FG02-04ER46134 and, in part, by the
MRSEC Program of the National Science Foundation under award
number DMR 02-13282.

%\begin{verbatim}

%\end{verbatim}

\end{document}